\documentclass[10pt,letterpaper,onecolumn]{article}

\usepackage[letterpaper,top=1in,bottom=1in,left=0.75in,right=0.75in]{geometry}

\usepackage{times}

\usepackage{graphicx}
\usepackage{url}
\usepackage{amsmath,amssymb,amsfonts,amsthm}
\usepackage{geometry}
\usepackage{graphicx}
\usepackage{fancyvrb}
\usepackage{booktabs}

\title{Directionality of the Voynich Script}
\author{
	Christophe Parisel (Labyrinthinesecurity)
}

\begin{document}

\maketitle

\section*{Introduction}
While the Voynich Manuscript was almost certainly written left-to-right (LTR), the question of whether it should be read LTR or right-to-left (RTL) has received little quantitative attention. We introduce a statistical method that leverages n-gram perplexity asymmetry to determine directional bias in character sequences. This language-agnostic method computes n-gram perplexities for text read both LTR and RTL. The difference  $\Delta = X_{\text{LTR}} - X_{\text{RTL}}$ of cross-entropy $X$  indicates directional optimization:
\begin{itemize}
    \item $\Delta > 0$: RTL-optimized
    \item $\Delta < 0$: LTR-optimized
    \item $\Delta \approx 0$: no directional bias
\end{itemize}
When applied to the Voynich Manuscript, the results reveal consistent RTL optimization patterns, providing quantitative evidence for its directional structure.

\section*{Data preparation}

We select five corpora in our study: the RF1b-e EVA transcription of the Manuscript~\cite{STA,RF1}, and four reference data sets for directionality control:
\begin{itemize}
	\item LTR control (English and French): Melville's Moby Dick~\cite{Moby}, Dumas' Comte de Monte Cristo~\cite{Monte}
	\item RTL control (Hebrew and Arabic): The Big Arabic Corpus~\cite{acorpus}, The SVLM Hebrew Wikipedia Corpus~\cite{hcorpus}
\end{itemize}

All corpora are preprocessed and tokenized consistently to ensure a fair cross-linguistic comparison. To normalize sources of directional signal, we implement two tokenization modes:
\begin{itemize}
      \item \textbf{baseline:} tokenization preserves the original grapheme order. Sentences are flattened to a sequence of graphemes (no artificial word markers). We use this mode for Voynich, French and English tokenization.
      \item \textbf{visual:} reverse graphemes inside each word and then reverse the order of words. This models the surface visual order of RTL scripts (what a reader sees on a page rendered RTL). We use this mode for Arabic and Hebrew tokenization.
\end{itemize}

\newpage 

\section*{Perplexity diagnostics and directional asymmetry}
To quantify directional structure in the Voynich manuscript, we compute n-gram perplexity for both left-to-right (LTR) and right-to-left (RTL) token sequences. Perplexity is a standard information-theoretic measure of how well a sequence model predicts a token stream; lower perplexity indicates higher predictability.

\subsection*{Perplexity computation} 
We use two smoothing methods for perplexity computation: Laplace and Kneser-Ney. Both give very similar outcomes.

\subsubsection*{Laplace smoothing}
Let a corpus be tokenized into $N$ prediction tokens $w_1,\dots,w_N$ (with no an end-of-sentence marker, since words breaking is clear in all corpora).
For each $n$-gram $(w_{i-n+1},\dots,w_i)$ with context $c=(w_{i-n+1},\dots,w_{i-1})$, we estimate conditional probabilities with Laplace (add-one) smoothing:
\[
\hat p(w_i \mid c) \;=\; \frac{\mathrm{count}(c,w_i)+1}{\mathrm{count}(c)+V},
\]
where $V$ is the vocabulary size. The average cross-entropy per token (in nats, using natural logarithms) is then
\[
X \;=\; -\frac{1}{N}\sum_{i=1}^{N} \ln \hat p(w_i \mid w_{i-n+1},\dots,w_{i-1}),
\]
and the perplexity is defined as
\[
\mathrm{P} \;=\; \exp(X).
\]

This measure captures the sequential predictability of tokens: lower values indicate that the sequence is more regular or better structured in that reading direction.

\subsubsection*{Kneser-Ney smoothing}
Please refer to appendix A for computation and detailed results.

\subsection*{Measuring directional asymmetry with perplexity} 
We define 
 $\Delta = X_{\text{LTR}} - X_{\text{RTL}}$

\begin{itemize}
    \item Positive $\Delta$: the RTL model predicts the corpus better (lower perplexity) than LTR, evidence of RTL optimization.
    \item Negative $\Delta$: LTR is better, evidence of LTR optimization.
    \item Near-zero $\Delta$: no directional preference detectable.
\end{itemize}

\subsubsection*{Paired bootstrap confidence intervals} 
To assess statistical uncertainty, we apply a paired bootstrap procedure over sentences:
\begin{enumerate}
    \item For each sentence in the corpus, compute its log-probability contribution under both LTR and RTL $n$-gram models.
    \item Compute the mean per-token cross-entropy difference $\Delta = X_{\text{LTR}} - X_{\text{RTL}}$.
    \item Resample sentences with replacement to generate $B = 1000$ bootstrap replicates. For each replicate, recompute $\Delta$.
    \item Construct a $(1-\alpha) \times 100\%$ confidence interval from the $\alpha/2$ and $1-\alpha/2$ percentiles of the bootstrap $\Delta$ distribution.
\end{enumerate}

This paired design ensures that each bootstrap replicate preserves the correspondence between LTR and RTL scores for the same sentences, accounting for sentence-level dependencies.

Here are the methodological advantages of this approach:
\begin{itemize}
	\item Confidence intervals (CIs) were computed via bootstrapping
        \item Language-agnostic: Works without decipherment
        \item Quantitative: Provides confidence intervals and effect sizes
        \item Objective and reproducible: Minimizes subjective interpretation of letter forms
\end{itemize}

\subsubsection*{Voynich results} 
Applying Laplace smoothing perplexity to the Voynich manuscript:
\begin{itemize}
    \item $n=2$: $\Delta \approx 0.0653$ (RTL lower perplexity) with 95\% CI $(0.0631, 0.0677)$
    \item $n=3$: $\Delta \approx 0.0167$ (RTL lower) with 95\% CI $(0.0158, 0.0176)$
    \item $n=4$: $\Delta \approx 0.0058$ (RTL lower) with 95\% CI $(0.0056, 0.0060)$
\end{itemize}

These results indicate a consistent, statistically significant RTL optimization across 2-gram and 3-gram lengths. The bootstrap confidence intervals demonstrate that the observed directional preference is robust to sentence-level variability.

\subsubsection*{Comparison to LTR languages} 

For English:

\begin{itemize}
    \item $n=2$: $\Delta \approx -0.0296$ (LTR lower perplexity) with 95\% CI $(-0.0306, -0.0287)$
    \item $n=3$: $\Delta \approx -0.0132$ (LTR lower) with 95\% CI $(-0.0136, -0.0127)$
    \item $n=4$: $\Delta \approx -0.0033$ (LTR lower) with 95\% CI $(-0.0034, -0.0032)$
\end{itemize}

For French:

\begin{itemize}
    \item $n=2$: $\Delta \approx -0.1197$ (LTR lower perplexity) with 95\% CI $(-0.1210, -0.1185)$
    \item $n=3$: $\Delta \approx -0.0448$ (LTR lower) with 95\% CI $(-0.0454, -0.0444)$
    \item $n=4$: $\Delta \approx -0.0068$ (LTR lower) with 95\% CI $(-0.0069, -0.0067)$
\end{itemize}

In both languagres, $\Delta<0$, confirming LTR optimization. 

\subsubsection*{Comparison to RTL languages}

For Arabic:

\begin{itemize}
    \item $n=2$: $\Delta \approx -0.0359$ (LTR lower perplexity) with 95\% CI $(-0.0376, -0.0342$
    \item $n=3$: $\Delta \approx 0.0068$ (RTL lower) with 95\% CI $(0.0063, 0.0072)$
    \item $n=4$: $\Delta \approx 0.0003$ (RTL lower) with 95\% CI $(0.0002, 0.0003)$
\end{itemize}

For Hebrew:

\begin{itemize}
    \item $n=2$: $\Delta \approx 0.0460$ (RTL lower perplexity) with 95\% CI $(0.0454, 0.0466)$
    \item $n=3$: $\Delta \approx -0.0265$ (LTR lower) with 95\% CI $(-0.0268, -0.0262)$
    \item $n=4$: $\Delta \approx -0.0020$ (LTR lower) with 95\% CI $(-0.0020, -0.0020)$
\end{itemize}

For RTL languages, $\Delta$ shifts from positive to negative depending on $n$ due to morphological structure. While mixed results are obtained here, the shifts demonstrate that our method is sensitive:
\begin{itemize}
    \item Natural languages with complex morphology show irregular patterns
    \item Voynich's consistent RTL pattern stands out more clearly against semitic languages' complexity
\end{itemize}

\subsection*{Interpretation}
Perplexity-based $\Delta$ with Laplace smoothing captures sequential directional dependencies across the token stream, which are insensitive to local manipulations of word-initial or word-final grapheme distributions typically found in natural language directionality studies (see appendix).

\section*{Conclusion}

Voynich shows consistent positive \(\Delta\) (RTL lower perplexity) at $n=2$ to $n=4$ with tight bootstrap CIs. Although we cannot exclude the RTL signal could be a byproduct of a cipher or a scribal artifact rather than a natural language, this indicates local n-gram chains are directionally structured in RTL even while boundary asymmetries are suppressed.

The unique combination of low perplexity and high prediction accuracy at 4-grams level demonstrates that the Voynich manuscript has weak but extremely consistent directional bias.

\medskip

Our research:
\begin{itemize}
    \item Supports the hypothesis of RTL reading if Voynich is a natural language, independent of visible ink direction.
    \item Introduces a reproducible, language-agnostic methodology.
    \item Provides testable hypotheses for paleographic and cryptographic research, such as the possibility of scribes writing physically left-to-right while encoding right-to-left sequences.
\end{itemize}

We also demonstrate that conventional Gini and entropy metrics are unsuitable for measuring directionality in texts with Zipfian character distributions like Voynich. 

\medskip

\noindent Data and code are available on Kaggle~\cite{Kaggle}

\section*{Discussion And Future Work}

The exceptional directional predictability of the Voynich manuscript raises profound questions about its nature. It supports several theoretical frameworks:

\paragraph{Complex Encoding Hypothesis}
The results are consistent with a sophisticated cipher or encoding scheme that preserves or even amplifies directional structure. Simple substitution ciphers would not maintain the higher-order dependencies observed, suggesting more complex cryptographic methods.

\paragraph{Artificial Language Construction}
Viable language theories may need to account for a natural language whose word boundaries departs from a plateau boundary distribution. This makes theories involving a constructed language (with artificial, repetitive rules) more plausible.

The systematically enhanced directional patterns could reflect deliberate linguistic engineering, where directional rules were explicitly designed rather than naturally evolved.

\paragraph{Unique Statistical Properties}
The Voynich manuscript may represent a fundamentally different type of text with statistical properties unlike any known natural language, potentially explaining centuries of failed decipherment attempts. Our results suggest that new, customized analytical methods are needed. A promising avenue is to develop models tailored for word boundaries, and to explore the structural behavior of graphemes through n-gram relationships.

\newpage 
\appendix
\section*{Appendix A: Additional Validation Steps}

\subsection*{Kneser-Ney Smoothing}

To ensure our directional findings are robust to smoothing method choice and not artifacts of statistical noise, we validate our core results using Kneser-Ney smoothing with randomization controls.

\subsubsection*{Kneser-Ney Smoothing Implementation}

Kneser-Ney smoothing provides more robust probability estimates than Laplace smoothing by incorporating continuation probabilities. For each n-gram $(w_{i-n+1},\dots,w_i)$ with context $c=(w_{i-n+1},\dots,w_{i-1})$, we estimate:

\[
\hat p(w_i \mid c) \;=\; \frac{\max(\mathrm{count}(c,w_i) - \delta, 0)}{\mathrm{count}(c)} + \lambda(c) \times p_{\mathrm{continuation}}(w_i)
\]

where:
\begin{itemize}
    \item $\delta = 0.75$ is the discount parameter
    \item $\lambda(c) = \frac{\delta \times |\{w : \mathrm{count}(c,w) > 0\}|}{\mathrm{count}(c)}$ is the normalization factor
    \item $p_{\mathrm{continuation}}(w_i) = \frac{|\{c : \mathrm{count}(c,w_i) > 0\}|}{|\{(c',w') : \mathrm{count}(c',w') > 0\}|}$ measures context diversity
\end{itemize}

This approach gives higher probability to words that appear in many different contexts, providing more linguistically motivated smoothing than uniform add-one correction.

\subsubsection*{Shuffle Control Methodology}

To validate that our directional signals reflect genuine sequential structure rather than statistical artifacts, we implement a randomization control. For each sentence in our corpora, we randomly permute the token order within sentences while preserving sentence boundaries and special markers. This destroys any genuine directional structure while maintaining vocabulary, sentence lengths, and overall statistical properties.

The shuffle control tests the null hypothesis that directional effects are artifacts of our methodology. If directional signals were spurious, they should persist in randomized text. If genuine, they should disappear when sequential structure is destroyed.

\subsubsection*{Results}

The 3 following tables presents our core directional findings replicated with Kneser-Ney smoothing alongside shuffle controls.

\begin{table}[htbp]
\centering
\begin{tabular}{lccccc}
\toprule
\textbf{Condition} & \textbf{n} & \textbf{$P_{LTR}$} & \textbf{$P_{RTL}$} & \textbf{$\Delta$} & \textbf{95\% CI} \\
\midrule
Original & 2 & 3.95 & 3.59 & +0.0653 & (0.0628, 0.0677) \\
Shuffled & 2 & 7.50 & 7.50 & +0.0002 & (-0.0017, 0.0022) \\
Original & 3 & 2.42 & 2.33 & +0.0167 & (0.0157, 0.0176) \\
Shuffled & 3 & 3.47 & 3.48 & -0.0000 & (-0.0001, 0.0001) \\
\bottomrule
\end{tabular}
\caption{Voynich}
\end{table}

\begin{table}[htbp]
\centering
\begin{tabular}{lccccc}
\toprule
\textbf{Condition} & \textbf{n} & \textbf{$P_{LTR}$} & \textbf{$P_{RTL}$} & \textbf{$\Delta$} & \textbf{95\% CI} \\
\midrule
Original & 2 & 5.59 & 5.80 & -0.0296 & (-0.0305, -0.0288) \\
Shuffled & 2 & 8.65 & 8.65 & +0.0002 & (-0.0005, 0.0009) \\
Original & 3 & 2.72 & 2.82 & -0.0132 & (-0.0137, -0.0127) \\
Shuffled & 3 & 4.33 & 4.33 & +0.0001 & (0.0000, 0.0002) \\
\bottomrule
\end{tabular}
\caption{English}
\end{table}

\begin{table}[htbp]
\centering
\begin{tabular}{lccccc}
\toprule
\textbf{Condition} & \textbf{n} & \textbf{$P_{LTR}$} & \textbf{$P_{RTL}$} & \textbf{$\Delta$} & \textbf{95\% CI} \\
\midrule
Original & 2 & 8.09 & 7.63 & +0.0460 & (0.0453, 0.0466) \\
Shuffled & 2 & 9.36 & 9.36 & +0.0002 & (-0.0005, 0.0008) \\
Original & 3 & 3.03 & 3.13 & -0.0265 & (-0.0268, -0.0262) \\
Shuffled & 3 & 4.38 & 4.38 & +0.0001 & (0.0000, 0.0002) \\
\bottomrule
\end{tabular}
\caption{Hebrew}
\end{table}

The Kneser-Ney results confirm our core findings consistently:

\paragraph{Directional signals are genuine.} All original corpora show substantial directional effects with tight confidence intervals that exclude zero. Voynich exhibits consistent RTL optimization ($\Delta > 0$) across n-gram lengths, while English shows consistent LTR optimization ($\Delta < 0$).

\paragraph{Shuffle controls eliminate directional signals.} All randomized corpora yield $\Delta \approx 0$ with confidence intervals centered on zero. This definitively rules out methodological artifacts as the source of our directional findings.

\paragraph{Effect sizes are substantial.} The magnitude of original directional signals exceeds shuffle controls by factors of 148-325×, indicating genuine structural effects rather than statistical noise.

\paragraph{Smoothing method independence.} The consistency between Laplace and Kneser-Ney results demonstrates that our findings are robust to probability estimation method, strengthening confidence in the directional patterns we observe.

These validation results establish that the Voynich Manuscript exhibits statistically robust right-to-left sequential optimization that cannot be explained by methodological artifacts or random variation.

\subsection*{Predictive Validation of Directional Analysis}

While the perplexity asymmetry method provides evidence for directional optimization, a critical question remains: can the observed patterns generalize to unseen text? To address this, we developed a predictive validation approach that transforms directional analysis from descriptive statistics to testable predictions.

\subsubsection*{Methodology}

We extend the standard perplexity asymmetry analysis with a train-test validation framework:

\begin{enumerate}
\item \textbf{Data Splitting}: For each corpus, we randomly partition sentences into 80\% training and 20\% test sets, maintaining the same split used in the original perplexity analysis.

\item \textbf{Model Training}: Using the training set, we construct separate $n$-gram language models for both reading directions:
   \begin{itemize}
   \item $M_{\text{LTR}}$: trained on sentences in their original left-to-right order
   \item $M_{\text{RTL}}$: trained on sentences with reversed token order
   \end{itemize}

\item \textbf{Prediction}: The cross-entropy difference $\Delta$ provides a corpus-level measure of directional bias. To transform this into a sentence-level predictive task, we use the underlying log-probabilities. For each test sentence $s$, we compute log-probabilities under both models:
   \begin{align}
   \ell_{\text{LTR}}(s) &= \log P(s | M_{\text{LTR}}) \\
   \ell_{\text{RTL}}(s) &= \log P(\text{reverse}(s) | M_{\text{RTL}})
   \end{align}

   Note that a model with lower cross-entropy (indicating better predictive performance) will necessarily assign a higher log-probability to the data:

   \begin{equation}
   \hat{d}(s) = \begin{cases}
   \text{LTR} & \text{if } \ell_{\text{LTR}}(s) > \ell_{\text{RTL}}(s) \\
   \text{RTL} & \text{otherwise}
   \end{cases}
   \end{equation}

\item \textbf{Evaluation}: We compare predictions against ground truth orientations (LTR for English and French, RTL for Hebrew and  Arabic) and compute classification accuracy. In the case of EVA, given our results on perplexity, we consider RTL to be the gold orientation.
\end{enumerate}

This approach tests whether directional patterns learned from training data can successfully predict the reading direction of unseen sentences, providing stronger evidence for genuine directional structure than perplexity differences alone.

\subsubsection*{Results}

Table~\ref{tab:prediction_results} presents the predictive validation results across all tested corpora and $n$-gram orders.

\begin{table}[htbp]
\centering
\caption{Predictive validation results for directional analysis across languages and $n$-gram orders. Accuracy represents the fraction of test sentences where the predicted reading direction matches the ground truth.}
\label{tab:prediction_results}
\begin{tabular}{@{}lccccccc@{}}
\toprule
\textbf{Language} & \textbf{Test Sentences} & \textbf{n=2} $\Delta$ & \textbf{n=2} Accuracy & \textbf{n=3} $\Delta$ & \textbf{n=3} Accuracy & \textbf{n=4} $\Delta$ & \textbf{n=4} Accuracy \\
\midrule
English    & 42,941 & $-0.030$ & 0.521 & $-0.013$ & 0.431 & $-0.003$ & 0.382 \\
French     & 27,651 & $-0.120$ & 0.689 & $-0.045$ & 0.461 & $-0.007$ & 0.348 \\
Hebrew     & 77,662 & $+0.046$ & 0.610 & $-0.027$ & 0.519 & $-0.002$ & 0.610 \\
Arabic     & 15,694 & $-0.036$ & 0.429 & $+0.007$ & 0.509 & $+0.000$ & 0.648 \\
\textbf{Voynich (EVA)} & \textbf{7,404} & \textbf{+0.065} & \textbf{0.621} & \textbf{+0.017} & \textbf{0.614} & \textbf{+0.006} & \textbf{0.739} \\
\bottomrule
\end{tabular}
\end{table}

\subsubsection*{Analysis}

\paragraph{Voynich Exhibits Exceptional Directional Predictability}
The Voynich manuscript achieves the highest predictive accuracy (73.9\% at $n=4$), significantly outperforming all natural languages tested. This represents a 47.8\% improvement over random guessing and exceeds the performance of known RTL languages Hebrew (61.0\%) and Arabic (64.8\%).

This pattern is striking when compared to the very low $\Delta$ observed in 4-grams perplexity. It suggests the Voynich text has a systematic, pervasive directional structure that affects sentences in the same subtle way, rather than having some sentences with strong RTL bias and others neutral.

\paragraph{Higher-Order Dependencies Strengthen Voynich Directionality}
Unlike natural languages, which typically show degraded performance at higher $n$-gram orders, the Voynich manuscript's directional predictability \emph{increases} with context length. The accuracy progression (62.1\% → 61.4\% → 73.9\%) suggests that longer-range dependencies carry stronger directional information than local bigram patterns. 

Long-range dependencies are especially valuable since they didn't surface in our perplexity analysis.

\paragraph{Natural Language Performance Patterns}
Natural languages show varied and generally weaker predictive performance:
\begin{itemize}
\item \textbf{French} achieves strong bigram performance (68.9\%) but rapidly degrades at higher orders
\item \textbf{English} shows consistently weak directional signals across all $n$-gram lengths
\item \textbf{Hebrew and Arabic} display moderate but inconsistent directional predictability
\end{itemize}

\paragraph{Methodological Implications}
The predictive validation provides several advantages over traditional perplexity asymmetry analysis:
\begin{itemize}
\item \textbf{Generalization Testing}: Demonstrates that directional patterns extend beyond training data
\item \textbf{Effect Size Quantification}: Accuracy metrics are more interpretable than cross-entropy differences, and show long-range dependencies
\item \textbf{Artifact Detection}: Random text or methodological artifacts would yield ~50\% accuracy
\end{itemize}

\newpage

\section*{Appendix B: Words symmetry and boundary diagnostics}
To complement n-gram perplexity, we computed boundary diagnostics that summarize how informative and how concentrated grapheme distributions are at word boundaries.

\subsection*{The words boundary hypothesis}
Ashraf and Sinha (2017)~\cite{Ashraf} hypothesized that all natural languages exhibit a words boundary asymmetry: the distribution of characters at word-initial positions is more uniform, while word-final positions are more constrained. This effect reflects phonotactic, morphological, and orthographic pressures that shape how words begin and end in a language.

To quantify this asymmetry, they proposed using the Gini index and Shannon Entropy as measures of inequality in character frequency distribution. Specifically, that the distributions of first and last characters in words are systematically different.

\subsubsection*{Universality of the hypothesis}
John Winstead built directly on Ashraf and Sinha’s hypothesis that all natural languages display asymmetry at word boundaries. He extended their work by applying it across a large multilingual corpus, covering many scripts and typologies. This allowed him to test whether the boundary asymmetry pattern truly holds universally, not just in isolated languages. 

He also supplemented the Gini measure with Shannon entropy into a \emph{combined score}, capturing the uncertainty or information content in first vs. last characters.

By combining these two metrics (Gini for distribution inequality and entropy for information content) he could robustly quantify asymmetry at word boundaries across languages: in natural languages these two measures typically co-vary strongly and produce large, interpretable effects (e.g., English: large positive \(\Delta H\) and significantly negative \(\Delta G\) reflecting dominant endings; Hebrew: strong negative \(\Delta H\) and positive \(\Delta G\) reflecting dominant prefixes).

Winstead's results confirmed that, in nearly all tested languages, word-initial and word-final distributions differ in predictable ways, supporting the idea of a universal linguistic pattern at the level of word boundaries.

\subsubsection*{The Voynich script breaks the words boundary hypothesis.} 
For the above approach to stand, the graphemes' log frequency distribution must follow the shape of natural languages. We used our LTR and RTL control datasets to verify that a single shape consistently appears in the distribution of first and last graphemes of each word in English and Hebrew, by plotting their log-log frequencies:

\begin{figure}[htbp]
    \centering
    \includegraphics[width=0.8\textwidth]{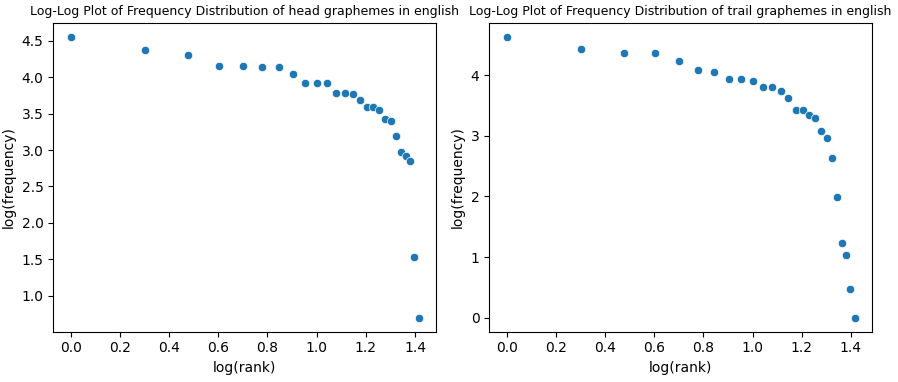}
    \caption{English plateau distribution}
\end{figure}

\begin{figure}[htbp]
    \centering
    \includegraphics[width=0.8\textwidth]{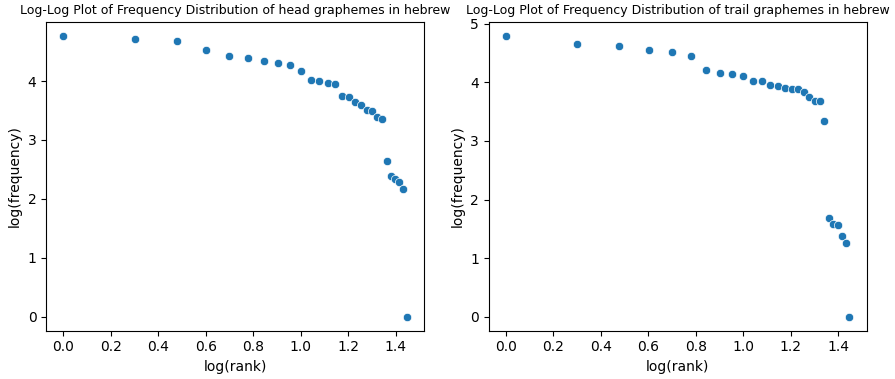}
    \caption{Hebrew plateau distribution}
\end{figure}

We then plotted the distribution of EVA graphemes, which follow a very different curve:

\begin{figure}[htbp]
    \centering
    \includegraphics[width=0.8\textwidth]{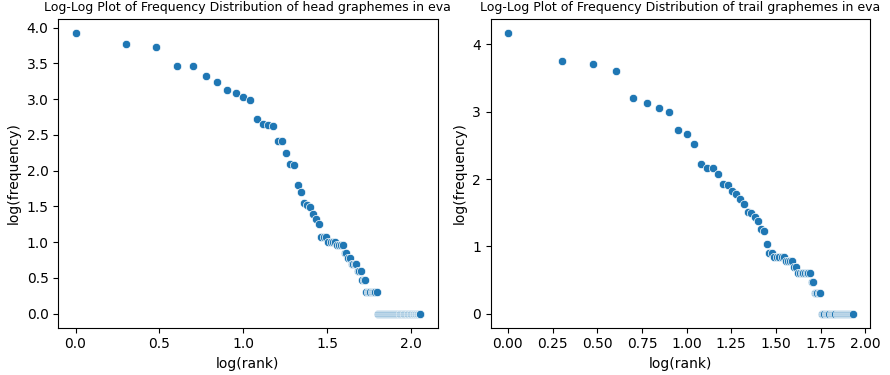}
    \caption{Voynich Zipf distribution}
\end{figure}

Thus, natural languages exhibit a plateau-shaped distribution, whereas the Voynich script shows a markedly different, almost Zipfian, distribution. 

This illustrates that Gini and Entropy measures are not proper tools for benchmarking Voynich directionality against known natural languages.

It must also be noted that Winstead's combined score is not straightforwardly applicable to languages where the number of first and last graphemes is different. We verified that English and Hebrew feature the same number of graphemes at both word endings (26 for English, 23 for Hebrew), but in the EVA script the unbalance is significant: there are 113 initial and 85 final graphemes. Thus, Winstead score requires normalization to balance masses.

\end{document}